\documentclass[12pt]{article}
\usepackage{amsmath,amssymb,cite}

\newtheorem{theorem}{Theorem}[section]   

\newtheorem{proposition}[theorem]{Proposition}

\newtheorem{remark}[theorem]{Remark}

\numberwithin{equation}{section}
\numberwithin{theorem}{section}

\newcommand{\bb}[1]{{\mathbb #1}}
\newcommand{\proof}{\noindent{\it Proof.} }
\newcommand{\qed}{\hfill $\square$\medskip}

\newcommand{\un}{\underline}

\newcommand{\address}[1]{\begin{itemize}\item[]\rm\raggedright #1 \end{itemize}}\def\eads#1{\address{E-mail: #1}}\def\mailto#1{{\tt #1}}\def\ams#1{\begin{itemize}\item[]\rm AMS classification scheme numbers: #1\par\end{itemize}} \newcommand{\rme}{\mathrm{e}}\newcommand{\rmd}{\mathrm{d}}                

\begin{document}

\title{Speedy motions of a body immersed in an infinitely extended medium}

\author{P. Butt\`a$^1$, G. Ferrari$^2$ and C. Marchioro$^1$}
\date{}
\maketitle 

\address{$^1$ Dipartimento di Matematica, SAPIENZA Universit\`a di Roma, P.le Aldo Moro 2, 00185 Roma, Italy}
\address{$^2$ Dipartimento di Matematica per le Decisioni Economiche, Finanziarie ed Assicurative, SAPIENZA Universit\`a di Roma, Via del Castro Laurenziano 9, 00161 Roma, Italy}
\eads{\mailto{butta@mat.uniroma1.it}, \mailto{giorgio.ferrari@uniroma1.it},  \mailto{marchior@mat.uniroma1.it}}
\ams{70F40, 70B05 , 70F45}

\begin{abstract} 
We study the motion of a classical point body of mass $M$, moving under the action of a constant force of intensity $E$ and immersed in a Vlasov fluid of free particles, interacting with the body via a bounded short range potential $\Psi$. We prove that if its initial velocity is large enough then the body escapes to infinity increasing its speed without any bound (\textit{runaway effect}). Moreover, the body asymptotically reaches a uniformly accelerated motion with acceleration $E/M$. We then discuss at a heuristic level the case in which $\Psi(r)$ diverges at short distances like $g r^{-\alpha}$, $g,\alpha>0$, by showing that the runaway effect still occurs if $\alpha<2$.
\end{abstract}


\section{Introduction}
\label{sec:1}

In the present paper we study the longtime behavior of a classical point body moving under the action of a constant force $\boldsymbol{E}$ and immersed in an infinitely extended medium. This problem has largely studied in the framework of kinetic theories, while here we want to obtain some rigorous results in the case of a fully Hamiltonian system.

Of course, when the medium is absent the body performs an uniformly accelerated motion. In the presence of a medium the motion is much more complicated and its asymptotic behavior depends on the nature of the background. In general, when the medium is homogeneous two different effects can happen: for strong body/medium interactions the motion of the body converges (in average) to an uniform one, while for weak body/medium interactions the body with a large initial velocity escapes to infinity increasing its velocity without bound. In the first case we have a reasonable model of viscous friction, in the second case we are in presence of a \textit{runaway particle} (see \cite{LL} for a discussion of this effect in the framework of kinetic theories).

Concerning the assumptions on the medium, a natural choice is to consider a system of infinitely many particles, mutually interacting via a pair potential $\Phi$, and interacting with the body via a potential $\Psi$. Unfortunately, this problem is very hard. Actually, the same definition    of the model is not obvious, due to the difficulty to prove the existence of dynamics for infinitely extended systems. We recall that the time evolution of systems with infinitely many particles has been investigated in several papers: we only quote some of the main results \cite {Lan68, Lan69, DoF77, FrD77, Fri85, BPY99, CMP00, CMS05}, other references can be found therein. Long time estimates drastically depend on the spatial dimensions of the system, and good enough estimates are known only for one dimensional models with bounded interactions \cite {CaM02}. The runaway effect has been shown to occur for large external force $\boldsymbol{E}$ when the medium is confined in a tube \cite{BCM03}, and for any strength of $\boldsymbol{E}$ in the case of particles moving along a line \cite{BCM04}. For singular interactions, the strict one-dimensional case becomes too particular, while in general we can only make conjectures. In \cite {BCM03}, for interaction singular as $r^{-\alpha}$ the threshold case is conjectured to be $\alpha=2$, but a rigorous proof seems too hard.

To make further steps we must simplify the medium. A possible way is to consider a mean field approximation, i.e.~to put the body in a Vlasov fluid. In the mean field approximation, well known in Astronomy and in Plasma physics, the background is composed by a gas of particles of mass going to zero and the number of particles per unit volume going to infinity in such a way that the mass density remains finite. The mean field limit of the dynamics of interacting particle systems in the case of  finite total mass has been studied in \cite {BrH77,Dob79,Neu81,Spo81} for bounded interaction and in \cite {HaJ07} for singular force like $r^{-\alpha}$, $\alpha < 1$. The limit in the case of infinite total mass has been proved in \cite{BCM02} in the case of bounded interaction and spatial dimension one. For unbounded mass, the existence of the dynamics has been proved in three dimensions of a Vlasov fluid with bounded interactions \cite {CCM01}, and in two dimension for the Helmholtz equation \cite {CMP02}.

In this framework some results have been recently obtained. In \cite{CMP06, Cav07} the interaction $\Psi$ is assumed of hard core form, precisely $\Psi (|\un{r}|) = \infty $ for $\un{r}\in \Lambda, \Psi=0$ otherwise, where $\Lambda$ is a convex set of $\bb R^3$ (this means that an element of the medium evolves freely out of $\Lambda$ and interacts elastically on $\partial \Lambda$). The initial phase space density of the medium is assumed spatially homogeneous out of $\Lambda$ and with Gaussian distribution of the velocities. It is proved the existence of a stationary state in the motion of the body for any intensity of a spatially constant force $\boldsymbol{E}$, and a detailed analysis of the approach to the limiting velocity is given. This result has been extended to forces not constant in space in \cite{CCM07}, and a similar result has been obtained  changing the elastic boundary condition with the diffusive ones (see \cite{ACM08,Aok08} for analytical and numerical results). In the case of singular interaction of the form $r^{-\alpha}, \alpha>0$ the problem is open. We only recall the paper \cite {BMM05}, where the runaway particle effect has been proved when $\Psi$ has the form $r^{-\alpha}$, $\alpha<2$, the medium is composed by a Vlasov fluid, the mutual interaction $\Phi$ is bounded, the influence of the body on the fluid motion is neglected, and the system has initially an one-dimensional symmetry.

In the present paper we assume $\Phi=0$, the body/medium interaction $\Psi$ bounded and with a short range, the background initially at rest, and we find not only the runaway effect but also the exact asymptotic velocity of the body. Then we discuss at an heuristic level the case in which $\Psi$ is always at short range but diverging at short distances as $r^{-\alpha}, \alpha<2$, case in which also the runaway effect is expected to take place. In \cite{BCM03} it has been conjectured that $\alpha =2$ is the threshold value for the existence or not of the runaway effect. A discussion on the existence of a stationary state for this model \cite{CaM09} improves this conjecture.

The plan of the paper is the following. In Section 2 we introduce the model and state the main result. In Section 3 we prove the result for $\Psi$ bounded. Finally, in Section 4 we discuss at a heuristic level the case of singular interaction. 

\section{The model}
\label{sec:2}

We consider the motion of a point body of mass $M$ and position $\boldsymbol{\xi}$, under the action of a constant force $\boldsymbol{E}$ of intensity $E$ and directed along the $x$-axis, i.e.~$\boldsymbol{E} = (E,0,0)$. The body is immersed in a fluid of free particles, that interact with the body via a force of pair potential $\Psi(|\boldsymbol{r}|)$. We assume the medium be a $Vlasov \ fluid$ of free particles in three dimensions, namely the pair of functions,
\begin{equation}
\label{p2}
(\boldsymbol{x},\boldsymbol{v}) \to (\boldsymbol{X}(\boldsymbol{x},\boldsymbol{v};t),\boldsymbol{V}(\boldsymbol{x},\boldsymbol{v};t)), \quad f_0(\boldsymbol{x},\boldsymbol{v}) \to f(\boldsymbol{x},\boldsymbol{v};t),
\end{equation}
with $(\boldsymbol{x},\boldsymbol{v},t)\in\bb R^3\times\bb R^3\times \bb R$, solution to
\begin{equation}
\label{p3}
\dot{\boldsymbol{X}}(\boldsymbol{x},\boldsymbol{v};t)=\boldsymbol{V}(\boldsymbol{x},\boldsymbol{v};t), 
\end{equation}
\begin{equation}
\label{p4}
\dot{\boldsymbol{V}}(\boldsymbol{x},\boldsymbol{v};t)= -\nabla\Psi(|\boldsymbol{X}(\boldsymbol{x},\boldsymbol{v};t)- \boldsymbol{\xi}(t)|),
\end{equation}
\begin{equation}
\label{p5}
(\boldsymbol{X}(\boldsymbol{x},\boldsymbol{v};0),\boldsymbol{V}(\boldsymbol{x},\boldsymbol{v};0))=(\boldsymbol{x},\boldsymbol{v}), 
\end{equation}
\begin{equation}
\label{p6}
f(\boldsymbol{X}(\boldsymbol{x},\boldsymbol{v};t),\boldsymbol{V}(\boldsymbol{x},\boldsymbol{v};t),t)=f_0(\boldsymbol{x},\boldsymbol{v}), 
\end{equation}
where $\boldsymbol{\xi}(t)$ is the coordinate of the point body, $f$ is the mass density in phase space of the free gas, and $(\boldsymbol{X},\boldsymbol{V})$ are the characteristics of the fluid. 

In the case of smooth initial data, $f(\boldsymbol{X}(\boldsymbol{x},\boldsymbol{v};t),\boldsymbol{V}(\boldsymbol{x},\boldsymbol{v};t),t)$  satisfies the differential equation,
\begin{equation}
\label{p7}
(\partial_t+\boldsymbol{v} \cdot \nabla_{\boldsymbol{x}}-\nabla\Psi(|\boldsymbol{x}-\boldsymbol{\xi}(t)|)\cdot \nabla_{\boldsymbol{v}})f(\boldsymbol{x},\boldsymbol{v};t)=0. 
\end{equation}
The system \eqref{p3}-\eqref{p6} is completed by the equation of motion of the body,
\begin{equation}
\label{p8}
M\ddot {\boldsymbol{\xi}}(t)= \boldsymbol{E} - \int\! \rmd\boldsymbol{x}\,\rmd\boldsymbol{v}\; \nabla_{\boldsymbol{\xi}} \Psi(|\boldsymbol{\xi}(t)-\boldsymbol{x}|)) \, f(\boldsymbol{x},\boldsymbol{v};t),
\end{equation}
with boundary condition
\begin{equation}
\label{p9}
\boldsymbol{\xi}(0) = \boldsymbol{\xi}_0, \quad \dot{\boldsymbol{\xi}}(0)= \dot{\boldsymbol{\xi}}_0.
\end{equation}

This kind of system, in which a Vlasov fluid is coupled to a massive body,
has been firstly introduced in connection with the so-called piston problem, see \cite{GrP99} and also \cite {LPS00} and references quoted therein. In our problem the fluid is unbounded and so, in principle, the existence of the solutions is not obvious. In fact, it is easy to exhibit  initial conditions for which the time evolution produces singularity in the motion after a finite time. This happens because very far away particles could arrive very fastly close to the body. These situations are pathologic from a physical point of view and can be removed in the case of bounded interaction by assuming that initially 
\begin{equation}
\label{p10}
f_0(\boldsymbol{x},\boldsymbol{v}) \leq \rho \left(\frac{\beta}{2\pi}\right)^{3/2}\rme^{-\beta\boldsymbol{v}^2/2}, \quad  \rho>0,  
\end{equation}
where $\beta=(kT)^{-1} >0$, $T$ is the temperature, and $k$ the Boltzmann constant, i.e.~$f$ is bounded by a homogeneous Gibbs state. In this case the proof of the existence of the solutions is quite easy.

\section{Bounded interaction}
\label{sec:3}

Assume $\Psi(|\boldsymbol{r}|)$ be a twice differentiable function of $\boldsymbol{r}\in \bb R^3$ and let  exist a positive constant $r_0<\infty$ such that $\Psi(r)=0$ if $r>r_0$. We assume that initially the point body is in the origin, $\boldsymbol{\xi}_0=(0,0,0)$, with velocity along the $x$-axis, $\dot{\boldsymbol{\xi}}_0=(\dot\xi_0,0,0)$, $\dot\xi_0>0$, and the fluid is at rest with constant density $\rho(\boldsymbol{x},0)=\rho_0>0$. By symmetry the body will move along the $x$-axis, i.e.~$\boldsymbol{\xi}(t) = (\xi(t),0,0)$ for some $\xi(t)\in\bb R$.
\begin{theorem}
\label{theor1}
For each intensity $E$ of the force, there exists a threshold value $\dot{\xi}_*$ such that for $\dot{\xi}_0>\dot{\xi}_*$ the body escapes to infinity with (asymptotically) a uniformly accelerated motion,
\begin{equation}
\label{3.1}
\lim_{t \to \infty} \frac{\dot{\xi}(t)}{t}= \frac{E}{M}.
\end{equation}
\end{theorem}
\proof We sketch the proof by emphasizing the main steps. We need to prove that the friction force exerted on the body by the fluid is initially  bounded and vanishes asymptotically in time. For this purpose we study the time evolution of a particle (i.e.~a characteristic) of the fluid.

First, we remark that if $\dot{\xi}(t)$ is large enough then each fluid particle interacts with the body at most one time. To prove this property we argue in the (non-inertial) reference frame where the body is at rest.  During the collision, the fluid particle of coordinates $\boldsymbol{x}(t)=(x_1(t),x_2(t),x_3(t))$ feels two forces, $-\nabla\Psi$ and $-\ddot{\boldsymbol{\xi}}(t) = (-\ddot{\xi}(t),0,0)$.
Assume for the moment that
\begin{equation}
\label{3.2}
\sup_{t\geq 0} |\ddot \xi (t)| < A < \infty,
\end{equation}
and
\begin{equation}
\label{3.3}
\dot \xi (t) > \frac{\dot \xi_0}{2} \quad \forall\, t \ge 0.
\end{equation}
(In the sequel, these properties will be proved to be verified for $A$ and $\dot\xi_0$ large enough).

Denote by $\tau$ the time at which the interaction begins and by $\tau+\delta$ the time at which it finishes. At time $\tau$ the velocity of the fluid particle is $\dot{\boldsymbol{x}}(\tau)=-\dot{\boldsymbol{\xi}}(\tau)$ and during the interaction its absolute value decreases at most by $(B+A)\delta$, where $B=\sup_r |\Psi'(r)|$. Hence
\begin{equation}
\label{3.4}
\delta \leq \frac{4r_0}{|\dot{\boldsymbol{\xi}}(\tau)|-2\delta (A+B)},
\end{equation}
that for large $\dot{\xi}_0$ implies
\begin{equation}
\label{3.5}
\delta \leq \frac{5r_0}{|\dot{\boldsymbol{\xi}}(\tau)|}.
\end{equation}
Hence, for large $\dot{\xi}_0$ the motion of the fluid particle does not change its direction and the particle escapes to $-\infty$.

This discussion suggests also the proof of the assumption \eqref{3.2}: the body could interact only with the fluid particles that are present in a cylinder of finite length and so the acceleration of the body is bounded uniformly in time. Hence \eqref{3.2} holds for $A$ large enough. 

Now, we come back to the reference frame in which the medium is initially at rest and we evaluate the momentum variation of the fluid particle during the collision with the body. Actually, a bound on $|\dot{\boldsymbol{x}}(\tau+\delta)|$ can be easily achieved by \eqref{3.5}, getting $|\dot{\boldsymbol{x}}(\tau+\delta)| \le  \textrm{const.}\, \|\Psi '\|_\infty \ |\dot{\boldsymbol{\xi}}(\tau)|^{-1}$, but we look for a more sophisticated bound useful in the next section. In fact, we expect that the momentum variation is really very small for large velocities: the effects produced when $x_1>\xi$ compensates in part those produced when $x_1<\xi$. To show this fact we introduce the quantity $\boldsymbol{p}(t) = (p_1(t),p_2(t),p_3(t))$ defined as
\begin{equation}
\label{3.6}
\boldsymbol{p}(t) = \dot{\boldsymbol{x}}(t)+ \frac{\Psi(|\boldsymbol{x}(t)-\boldsymbol{\xi}(t)|)}{|\dot{\boldsymbol{x}}(t)-\dot{\boldsymbol{\xi}}(t)|^2} (\dot{\boldsymbol{x}}(t)-\dot{\boldsymbol{\xi}}(t)).
\end{equation}
Obviously, before and after the collision,
\begin{equation}
\label{3.7}
\boldsymbol{p}(t) = \dot{\boldsymbol{x}}(t).
\end{equation}
We shall show that during the collision $\dot{\boldsymbol{p}}(t)$ is quite small and consequently the momentum variation is small. Recall $[\tau,\tau+\delta]$ denotes the time interval during which the collision occurs. By \eqref{3.7} and remembering that before the collision the fluid particle is at rest, we have,
\begin{equation}
\label{3.8}
\dot{\boldsymbol{x}}(\tau+\delta) = \int_{\tau }^{\tau+ \delta}\!\rmd s\; \dot{\boldsymbol{p}}(s),
\end{equation}
which implies
\begin{equation}
\label{3.9}
|\dot{x}_1(\tau+\delta)|\le \int_{\tau }^{\tau+\delta}\!\rmd s\;  |\dot  p_1(s)|,
\end{equation}
\begin{equation}
\label{3.10}
|\dot{x}_2(\tau+\delta)|\le \int_{\tau}^{\tau + \delta}\!\rmd s\;  |\dot p_2(s)|,
\end{equation}
\begin{equation}
\label{3.11}
|\dot{x}_3(\tau+\delta )|\le \int_{\tau}^{\tau + \delta}\!\rmd s\;  |\dot p_3(s)|.
\end{equation}
We next use these equations to obtain finer estimates on $|\dot{\boldsymbol{x}}(\tau+\delta)|$. From now on $C$ denotes a generic positive constant whose numerical value may change from line to line.  
We start by studying $|\dot{x}_2(\tau+\delta)|$.  From \eqref{3.6} we have,
\begin{eqnarray}
\label{3.12}
\dot{p}_2(t)\,&=&\,\ddot{x}_2 \left[1 + \frac{\Psi(|\boldsymbol{x}-\boldsymbol{\xi}|)}{|\dot{\boldsymbol{x}}-\dot{\boldsymbol{\xi}}|^2}\right] + \dot{x}_2 \frac{\Psi'(|\boldsymbol{x}-\boldsymbol{\xi}|)\;(\boldsymbol{x}-\boldsymbol{\xi})\cdot(\dot{\boldsymbol{x}}-\dot{\boldsymbol{\xi}})}{|\boldsymbol{x}-\boldsymbol{\xi}|\,|\dot{\boldsymbol{x}}-\dot{\boldsymbol{\xi}}|^2}\,+\, \nonumber \\
\,&-&\,\frac{2\,\dot{x}_2\,\Psi(|\boldsymbol{x}-\boldsymbol{\xi}|)\,(\dot{\boldsymbol{x}}-\dot{\boldsymbol{\xi}})\cdot(\ddot{\boldsymbol{x}}-\ddot{\boldsymbol{\xi}})}{|\dot{\boldsymbol{x}}-\dot{\boldsymbol{\xi}}|^4}.
\end{eqnarray}
We observe that $|\ddot{x}_2| < |\Psi '(|\boldsymbol{x}-\boldsymbol{\xi}|)|  $ (by the equations of motion) and, for $\dot \xi_0$ large enough, $\dot \xi > \frac 12 |\dot{\boldsymbol{x}}-\dot{\boldsymbol{\xi}}|$. By \eqref{3.10},
\begin{equation}
\label{3.13}
|\dot{x}_2(\tau+ \delta)|\le  C \int_{\tau}^{\tau + \delta}\!\rmd s\; \big|\Psi'(|\boldsymbol{x}(s)-\boldsymbol{\xi}(s)|)\big| \, \big[1+\|\Psi\|_\infty \dot\xi(s)^{-2} \big].
\end{equation}
Changing variable $\rmd s = \dot \xi^{-1} \rmd \xi$ into the last integral 
and observing that 
$$
\int_{\tau}^{\tau + \delta}\!\rmd s\; \Psi'(|\boldsymbol{x}(s)-\boldsymbol{\xi}(s)|) \le C \, \|\Psi\|_\infty \, \zeta^{-1}, 
$$
where
\begin{equation}
\label{3.14}
\zeta = \min_{\tau\le s\le \tau+\delta} \dot \xi(s)^{-1},
\end{equation}
we have
\begin{equation}
\label{3.15}
|\dot{x}_2(\tau+\delta)|\le  C \, \frac{\|\Psi\|_\infty}{\zeta}\left(1 + \frac{\|\Psi\|_\infty}{\zeta^2}\right).
\end{equation}
Due the cylindrical symmetry the same estimate holds for $|\dot{x}_3(\tau+\delta)|$. We then turn to the estimate of $|\dot{x}_1(\tau+\delta)|$. By \eqref{3.6} and the equations of motion,
\begin{eqnarray}
\label{3.16}
\dot{p}_1(t) & = & - \frac{ \Psi'(|\boldsymbol{x} - \boldsymbol{\xi}|)}{|\boldsymbol{x} - \boldsymbol{\xi}||\dot{\boldsymbol{x}}-\dot{\boldsymbol{\xi}}|^2} \left[(x_1-\xi)(\dot{x}_2^2  + \dot{x}_3^2) + (\dot{\xi}-\dot{x}_1)(x_2\dot{x}_2 +x_3\dot{x}_3)\right] \nonumber \\ && +   \frac{ \Psi(|\boldsymbol{x} - \boldsymbol{\xi}|)}{|\dot{\boldsymbol{x}}-\dot{\boldsymbol{\xi}}|^2}\left[-(\ddot{\xi}-\ddot{x}_1) +\frac{ 2(\dot{\xi}-\dot{x}_1)(\dot{\boldsymbol{\xi}}-\dot{\boldsymbol{x}})\cdot (\ddot{\boldsymbol{\xi}}-\ddot{\boldsymbol{x}})}{|\dot{\boldsymbol{x}}-\dot{\boldsymbol{\xi}}|^2}\right].
\end{eqnarray}
By plugging the bound \eqref{3.15} and again equations of motion into \eqref{3.16}, by \eqref{3.9} and recalling the definition \eqref{3.14} we obtain
\begin{equation}
\label{3.17}
|\dot{x}_1(\tau+\delta)| \le  C\, \frac{\|\Psi\|_\infty^2}{\zeta^3}
\left(1+\frac{\|\Psi\|_\infty}{\zeta^2} + \frac{\|\Psi\|_\infty^2}{\zeta^4}+ \frac{\|\Psi\|_\infty^3}{\zeta^6}\right).
\end{equation}
Hence, since $\|\Psi\|_\infty < \infty$, each fluid particle increases the $x$-component of its momentum by a quantity at most proportional to $\zeta^{-3}$. Since the body meets per unit time a number of fluid particles proportional to $\dot{\xi}(t)$, it looses per unit time an amount of momentum bounded as below,
\begin{equation}
\label{3.19}
\textrm{[lost momentum per unit time]} \le C \, \dot{\xi}(t)^{-2}.
\end{equation}

The other steps in the proof are easy: we show that, for $C_1,\dot{\xi}_0>0$ large enough,  
\begin{equation}
\label{3.20}
M \dot\xi(t) \ge \frac{3M}4 \dot\xi_0 + E\, t - \int_0^t\!\rmd s\, \frac{C_1}{\dot\xi(s)}.
\end{equation}
Indeed, denote by $t^*$ the supremum of the times $T$ such that the inequality \eqref{3.20} is verified for any $t\in [0,T]$. Since $M\dot \xi(0) = M\dot\xi_0$, we have $t^*>0$ by continuity. We decompose the time interval $[0,t^*)$ into a partition of intervals $\delta_k$ of length of the order $\dot\phi\Big(\sum_{i=1}^k \delta_i\Big)$, where $\phi(t)$ is the solution to the integral equation
$$
M  \dot\phi(t) = \frac{3M}4 \dot\xi_0 + E\, t - \int_0^t\!\rmd s\, \frac{C_1}{\dot\phi(s)}.
$$
Since \eqref{3.20} suddenly implies the assumption \eqref{3.3} for $\dot\xi_0$ large enough, the bound \eqref{3.19} is valid in each interval $\delta_k$. It follows that the approach to the asymptotic (uniformly accelerated) motion of $\xi(t)$ is faster than that of $\phi(t)$, whence $t^*=+\infty$ for large enough $C_1$ and $\dot\xi_0$. So \eqref{3.20} is valid at any positive time and the theorem is proved. 
\qed
\begin{remark}\rm
We have assumed the density of the fluid constant in the whole space. Of course, the result and the proof do not change if we assume the initial density $\rho_0(\boldsymbol{x})$ with an axial symmetry around the $x$-axis and $\rho_0(\boldsymbol{x}) \to \textrm{const.}$ fast enough as $|\boldsymbol{x}| \to \infty$. Indeed, this only changes the transient evolution, but the asymptotic behavior of the motion remains unchanged.

It is natural to ask what happens if we assume the fluid initially distributed as a Gibbs state like \eqref{p10} (Theorem~\ref{theor1} here corresponds to the case $\beta =\infty$). For $E$ larger than $|\Psi'|$ the result is still true \cite{Sga08}, while for smaller $E$ it depends on the sign of $\Psi'(r)$: we conjecture that if $\Psi'(r)\ge 0$ the result remains valid; while, if $\Psi'(r)<0$ for some $r$, the quantity $\dot\xi(t)/t$ should converge to some constant possibly smaller than $E/M$. In fact, the body could capture a small part of fluid, which remains with it forever, with the only effect to increase the effective inertia (i.e.~mass) of the body.
\end{remark}

\section{Singular interaction}
\label{sec:4}

The spirit of this section is different from the previous one, where we have given a mathematically rigorous proof on the asymptotic behavior of the system. Here we confine ourself to a heuristic discussion of the main issues in a possible proof.

We assume $\Psi(r)$ be a twice differentiable function for $r>0$ and there exist two positive constants $r_1,r_0<\infty$ such that $\Psi=gr^{-\alpha},g>0,\alpha>0$ if $r< r_1$,  $\Psi$ equals to a not increasing function if $r_1\leq r \leq r_0$, $\Psi=0$ if $r>r_0$. Initially we  put the point body in the origin with a fix velocity $\xi(0)=\xi_0>0$, and the fluid at rest with a constant density  $\rho(\boldsymbol{x},0)=\rho_0>0$. We discuss the following proposition, analogous to Theorem~\ref{theor1}.
\begin{proposition}
\label{prop1}
For each intensity $E$ of the force, there exists a threshold value $\dot{\xi}_*$ such that for $\dot{\xi}_0>\dot{\xi}_*$ the body escapes to infinity with (asymptotically) a uniformly accelerated motion,
\begin{equation}
\label{4.0}
\lim_{t \to \infty} \frac{\dot{\xi}(t)}{t}= \frac{E}{M}.
\end{equation}
\end{proposition}
The strategy of the proof is similar to that used in the bounded case: we assume a behavior of the body and we evaluate the viscous friction produced by the fluid. Using an adiabatic invariant we show that the friction is very small and vanishes at long times, so that the behavior of the body is better than the assumption. Unfortunately, we do not prove some (minor) steps and the discussion remains only heuristic.

In the present analysis two new points arises: $a)$ the interaction is singular and so the particles of fluid near the $x$-axis drastically change their velocity whatever large is the velocity of the body; $b)$ some particles of the fluid can collide infinitely many times. The first difficulty is solved by proving the these \textit{bad} particles are very few when the velocity of the body becomes large; the second difficulty is solved by showing the the effect of recollisions is negligible.

As in the bounded case we assume that
\begin{equation}
\label{4.2}
\sup_{t\geq 0} |\ddot \xi (t)| < A < \infty .
\end{equation}
We do not prove it rigorously. However physically it is quite obvious; moreover, at the end of the discussion it will be clear that a proof using this assumption should imply that the acceleration goes to zero in average on a time interval of length vanishing as $t \to \infty$. But a rigorous proof of assumption \eqref{4.2} appears cumbersome and not trivial.

We explicitly discuss the more difficult case $\alpha >1$, starting with  the study of the first collision.

We analyze the motion of a fluid particle. We denote by $\eta$ its impact parameter, that is the distance between the incoming fluid particle and the $x$-axis. In the plane $(x_2,x_3)$ we consider the disks $D_k$, centered in the origin and of radius $\eta_k = 2^k\eta_0 $, $k=0,1,2,\ldots$ We choose $\eta_0=\dot{\xi}(t)^{- (1+\epsilon)}$, $\epsilon>0$, so that the particles with $\eta\le\eta_0$ contribute at most as $C\dot \xi(t)^{-2\epsilon}$ in the momentum transferred to the body (per unit time). In fact, the area of this disk is $\pi\eta_0^2$, the body meets per unit time a number of fluid particles proportional to $\dot \xi (t)$, in each collision it looses a momentum $2\dot{\xi}(t)$ (corresponding to an elastic collision with a disk) and so the lost momentum due to particle hitting in this circle is at most $C \dot \xi^2 (t) \eta_0^2 = C \dot \xi(t)^{-2\epsilon}$.
 
We now evaluate the effect of the collisions in the $k$-annulus $D_k\setminus D_{k-1}$. First we remark that, by arguing as at the beginning of the previous section, it is possible to show that the body has only one collision with each fluid particle posed in a $k$-annulus, and then it overcomes such particles. Hence, the lost momentum is given by \eqref{3.17}, where now $\|\Psi \|_\infty = C \, (\eta_k)^{-\alpha} $ and the dominant term is $\|\Psi \|_\infty^2 \dot\xi(t)^{-3}$ (here we approximate $\zeta$, see \eqref{3.14}, by $\dot\xi(t)$). The area $A_k$ of the $k$-annulus is $C\eta_k^2$ and the intensity is of the order of $\dot \xi(t)$. In conclusion, the contribution $I$ to the lost momentum due to the fluid particles posed in all the annuli is bounded as
 \begin{equation}
\label{4.3}
I \le C \, \dot\xi(t)^{[(2\alpha-2)(1+\epsilon)-2]} \sum_{k=1}^{k^*} 2^{2k(-\alpha +1)}, \qquad \eta_{k^*}\ge r_0.
\end{equation}
If we choose $\epsilon < (2/\alpha)-1$ we obtain 
\begin{equation}
\label{4.4}
I\le C\, \dot{\xi}(t)^{-2\epsilon},
\end{equation}
that is a bound analogous to the one obtained for the first disk.

We now roughly discuss the effect of multiple collisions. For the moment, we suppose that the body is governed by a uniformly accelerated motion. The only possibility for a fluid particle to hit the body many times is to be quite close to the $x$-axis. In particular, we start by evaluating how much close for hitting the body twice. Denote by $\sigma$ the impact parameter of the fluid particle and let $t_1$ be the time at which the first collision occurs. Due to the monotonicity of the potential, during the collision the fluid particle feels a repulsive force from the $x$-axis which varies its orthogonal velocity by a quantity
\begin{equation}
\label{4.5}
\delta v^\perp \simeq \int_{t_1}^{t_1+\delta}\!\rmd s\, |\nabla \Psi (\boldsymbol{x}(s)-\boldsymbol{\xi}(s))| \, \frac{\sigma}{r_0}.
\end{equation}
Changing variable $\rmd s = \dot \xi^{-1} \rmd \xi$ we obtain
\begin{equation}
\label{4.6}
\delta v^\perp \gtrsim C \, \frac{\sigma}{r_0} \dot{\xi}(t_1)^{-1} r_m^{-\alpha},
\end{equation}
where $r_m$ is the minimal distance between the fluid particle and the body. Since $r_m \approx C \, \dot{\xi}^{-2/\alpha}$ we conclude that
\begin{equation}
\label{4.61}
\delta v^\perp \gtrsim C \, \sigma \dot{\xi}(t_1).
\end{equation}
After the collision, the fluid particle has gained a velocity which is double than the speed of the body. But the particle then preserves its velocity while the body accelerates almost uniformly, and so it can be reached again after a time of the order of $\dot \xi (t_1)$. In conclusion, the fluid particle can hit the body twice if its impact parameter $\sigma$ is very small,
\begin{equation}
\label{4.9}
\sigma \lesssim C \, \dot \xi(t_1)^{-2}.
\end{equation}
Actually, we are interested in the case when the fluid particle has a third collision. For this to happen the impact parameter has to be very small, precisely
\begin{equation}
\label{4.10}
\sigma \lesssim C\, \dot\xi(t_1)^{-2} \dot\xi(t_2)^{-2} \le C\, \dot\xi(t_1)^{-4}
\end{equation}
where $t_2$ is the time of the second collision and we used $\dot\xi(t_2)\ge \dot\xi(t_1)$.

To apply these ideas to the true case, some modifications are needed: taking into account that the motion of the body is not exactly uniformly accelerated, but with a small perturbation; moreover, to evaluate the effect of multiple collisions at a fixed time, we should go back in the past up to the initial time; finally, for the repulsive force we need a lower bound (not an approximation as in previous discussion). We do not perform this analysis in detail, but we hope to have convinced the reader that the effects of multiple collisions are negligible.

The further steps are similar to that of the bounded case: we assume that 
\begin{equation}
\label{4.11}
M \dot{\xi}(t) \ge  \frac{3M}4 \dot{\xi}_0 + E t - \int_0^t\!\rmd s\, \frac{C}{\dot\xi(s)^{\epsilon}},
\end{equation}
and by \eqref{4.4} we prove that the approach is faster. Then the previous bound holds at any positive time and the body motion converge to a uniform accelerated one as $t\to \infty$.

\section*{Acknowledgments}
Work performed under the auspices of GNFM-INDAM and the Italian Ministry of the University (MIUR).

\small

\end{document}